\documentstyle[aps,psfig,epsf]{revtex}
\author{
Antonio Coniglio$^a$ and Mario Nicodemi$^{a,b}$ 
}
\address{
\vspace{0.2cm}
$^a$ Dipartimento di Fisica, Universit\'a di Napoli ``Federico II'',
INFM and INFN di Napoli, Via Cintia, 80126 Napoli, Italy\\
$^b$ Department of Mathematics, 
Imperial College, London, SW7 2BZ, U.K.
}

\title{A Statistical Mechanics Approach to the Inherent States of Granular 
Media}

\newcommand{\lan}{\langle}
\newcommand{\ran}{\rangle}

\begin{document}

\maketitle
\bigskip
\begin{abstract}
We consider a Statistical Mechanics   approach to granular systems by
following the original ideas developed by Edwards.   We use the
concept of ``inherent states'', defined as the stable configurations in the 
potential energy landscape, introduced in the context of glasses. 
Under simplifying assumptions, the equilibrium inherent states 
can be characterized by a configurational temperature, $1/\beta$. 
We link $\beta$ to Edwards' compactivity and address the problem of 
its experimental measure. We also discuss the possibility to describe 
the time dependent distribution probability in the inherent states 
with an appropriate master equation.
\end{abstract} 

\bigskip

The possibility to describe general features in the physics of
granular   media with the concepts of Statistical Mechanics was
suggested few   years ago by Edwards \cite{Edwards}. 
As much as systems of standard   Statistical Mechanics, each 
macroscopic state of a granular medium corresponds 
to a huge number of microstates.  
Furthermore, these systems show very general reproducible macroscopic
behaviors characterized by a few control parameters \cite{JNBHM}.     
In granular media, however, in absence of   some external driving
the microstates are ``frozen'', i.e., they don't
evolve in time as thermal energy is negligible compared
to   gravitational energy. Therefore the external thermal bath can be 
considered at zero temperature.
Thermal motion of grains can be replaced, instead,
by agitation   induced by shaking or other form of driving,
which may generate a dynamics among ``frozen'' microstates. In facts, 
granular systems are just one example of a broad category   of
materials which can be found in ``frozen states''. An other
interesting example is the state of a supercooled  
liquid which has been cooled at zero temperature  and falls in one of
the microstates corresponding to the minima of the potential energy, called   
``inherent structures'' \cite{Stillinger}. 
Actually, the analogy between slow dynamics in granular media and glassy 
behaviors of supercooled liquids has been proposed in various forms 
\cite{Edwards,Knight,mehta1,CH,NCH,degennes,our_rev}. 

In this Letter we will further extend this analogy  to develop a statistical 
mechanics approach to granular materials, following Edwards  pioneering ideas 
\cite{barrat}. 
For simplicity we consider here a granular material as made 
of particles which may have a distribution of shapes and interacting 
via hard core potentials, so that they are not allowed to overlap. Due
to gravitational energy, each particle configuration $\{r_i\}$ corresponds
to a potential energy $U\{r_i\}$. The stable packing configurations
are the minima or saddle points of the potential energy 
or, more generally, all the states which, due for instance to friction, 
are mechanically stable. These states are the frozen configurations 
which by analogy with the glass terminology we call {\em inherent
states}. If there is no ambiguity throughout the paper we call them simply
states. We label each state by an index $s$ and call $E_s$ the
corresponding energy.


The dynamics from one state to another can be induced by some form of driving  
such as, for instance, by sequences of shakes or ``taps'' of given amplitude  
as in typical experimental realizations (see references \cite{JNBHM,Knight}). 
The kinetic energy driven in the packs by the ``shakes'' 
is totally dissipated when the shake is over and  
the system is almost instantaneously ``frozen'' in one  
of the inherent states.  
The actual dynamics depends on several parameters such as the tap amplitude  
$\Gamma=a/g$ ($a$ is the shake peak acceleration and $g$ the gravity  
acceleration), frequency and others \cite{nota1}.
We say that the system is at stationarity if its macroscopic properties 
don't change any longer with respect to the given dynamics 
allowing transitions from one microstate to another. 
We refer to these states as stationary states, or quasi-stationary if the
macroscopic properties change very slowly.  


A very important issue is to individuate the states distribution, 
namely what is the probability to find the system in a given inherent state 
$s$. We can expect that under stationary condition \cite{nota_qs} 
the system has been ``randomized'' enough, and 
therefore, following essentially Edwards original ideas, we make the
assumption  that  such a distribution is given by a maximization 
of the entropy under the condition that the average energy is fixed. 

More precisely 
we consider a statistical ensemble of equivalent systems all prepared
in the same way. 
We indicate with $\{E_r\}$ the energies of the accessible inherent microscopic 
states of each system and with $n_r$ the number of systems   
with energy equal to $E_r$. The average energy per system is thus  
$
E=\sum_r P_r E_r 
$
where  
$P_r=n_r/N$ is the probability to have a system in the inherent  state $r$.

The configurational entropy can be defined as $S=-\sum_r P_r \ln P_r$. 
We assume that the stationary distribution is given by the maximal
entropy under the constraint which fixes the average energy $E$. 
This requirement leads to the Gibbs distribution function:  
\begin{equation} 
P_r=\frac{e^{-\beta E_r}}{Z} 
\label{gibbs}
\end{equation} 
where the partition function $Z=\sum_r e^{-\beta E_r}$ is a normalization 
factor and  
$\beta$ a Lagrange multiplier determined by the constraint on the energy. 

As in standard Statistical Mechanics, we can show that in the thermodynamic
limit $S$ is the logarithm  
of the number of microscopic inherent states, $\Omega$, corresponding  
to the macroscopic energy $E$. 
Actually, we have that $\ln Z = S - \beta E$. 
On the other hand $Z$ can be written as 
$Z= \sum_r e^{-\beta E_r} =\sum_{\cal E} \Omega({\cal E}) e^{-\beta {\cal E}}$ 
and using the steepest descent method, in the thermodynamic limit one obtains: 
$\ln Z = \ln \Omega(E)-\beta E$. Thus, we have that:  
$S=\ln \Omega(E)$ 
and
\begin{equation}
\beta=\frac{\partial S}{\partial E}.
\label{beta}
\end{equation}
We call  $\beta^{-1} \equiv T_{conf}$
the {\em configurational temperature}. 
We note that beside the  configurational temperature $T_{conf}$ 
there is a second ``temperature'' related to the grains kinetic energy, 
the external effective bath temperature, $T_{bath}$, which is zero. 
While the fast degree of freedom, related
to the momentum distribution, equilibrates with the temperature of the
external bath resulting into a zero kinetic energy,  
the non zero configurational temperature characterizes the 
equilibrium distribution among the inherent states. 
Note that two granular systems in ``thermal contact'' have always 
the same $T_{bath}=0$ while 
the configurational temperature $T_{conf}$ may be different.
However
the configurational temperature 
characterizes the equilibrium states of the system 
and depends only on the average energy
of the inherent states and not on the particular dynamics used.
$T_{conf}$ enters as a parameter in the
equilibrium distribution allowing to substitute, as in usual Statistical 
Mechanics, time averages, which are theoretically difficult to evaluate, 
with ensemble averages. 

The existence of two effective temperatures is quite familiar in the context 
of glassy systems\cite{bouchaud,peliti}. 
If one quenches these systems in their glassy phase at a low bath temperature, 
$T_{bath}$, after a relative short transient the fast degree of freedom
come quickly in equilibrium with $T_{bath}$, while
the slow degree of freedom related to configurational rearrangement 
fall out of equilibrium and are characterized by an internal temperature
$T_{int}$ \cite{peliti,sciortino}. 
Interesting enough when $T_{bath}=0$, $T_{int}$ coincides with the configurational temperature 
introduced in eq.(\ref{beta}): $T_{int} = T_{conf}$ 
\cite{sciortino,franz}.

There are however some differences between  glassy 
systems quenched at $T_{bath}$ almost equal to zero  and granular materials.  
In the former the energy depends on the equilibrium temperature
before quenching and 
the dynamics among the inherent states is due to an
activated thermal process at extremely low temperature. In granular media 
the energy depends on the shaking amplitude $\Gamma$ and the dynamics
in general proceeds with  a tapping process with the same fixed amplitude. 
It would be interesting to study glassy materials using the dynamics of
granular materials and vice-versa.
For example one could introduce in glass formers a ``tap dynamics'', in which 
the temperature $T_{bath}$ is cyclically varied between zero and a finite 
value $T_{\Gamma}$ \cite{dean}. 
On the contrary in granular material one might first reach a given value of 
the energy by shaking the system with an amplitude $\Gamma$ and then evolve 
the system by gently shaking with very low value of $\Gamma$.


We now show that in the particular case in which the particles density 
$\rho$ is constant, $\beta$ can be easily related to the ``compactivity'',
introduced by Edwards in his seminal papers \cite{Edwards}. 
An ``inherent  state'' can be characterized by the 
particle density function $\rho(\vec r)$.  
When the system has a constant density profile, i.e., $\rho(\vec r)=\rho_0$ 
in the packs, a one to one correspondence exists between  energy and volume.  
Let's consider for sake of simplicity only hard core and gravity interactions. 
Actually, if  
$A$ is the horizontal area of the container and $h$ the packs height  
(its volume is $V=hA$), 
the system energy is: $E=mg\rho_0 A h^2/2={\cal P} V/2$, where  
$m$ is the grain mass and the characteristic pressure  
${\cal P}=mg\rho_0 h$ is the weight of the sample per unit surface.  
Then we have: 
\begin{equation}
\beta \equiv \frac{\partial S}{\partial E}=\frac{2}{{\cal P}} 
\frac{\partial S}{\partial V}=\frac{2}{{\cal P}} X_E^{-1} 
\end{equation}
The quantity $X_E^{-1}=\frac{\partial S}{\partial V}$  
is  Edwards' ``compactivity''\cite{Edwards}, which in this case is  
simply proportional to $\beta$.


To see how the formalism can work we consider 
a simple example where  $\rho$ is not a constant. 
As stated, the inherent states of a granular system can be characterized 
by their coarse-grained  
density spatial distribution, $\rho(\vec r)$, and the partition function  
of the system may be written as:  
$
Z=\int{\cal D}\rho ~ e^{-\beta{\cal F}[\rho]} 
$, 
where  
$
{\cal F}[\rho]=E[\rho]-\beta^{-1} S[\rho] 
$
is the free energy functional for the system in its inherent states 
and the integral is over all inherent states. In the simplest cases 
the  gravitational energy  $E[\rho]$ is given by:
$
E_g[\rho]=mg \int_V d{\vec r} ~ z\rho({\vec r}) 
$
($z$ is the component of $\vec r$ along gravity acceleration, $\vec g$).  
$S$ can be approximated  by the entropy of a lattice gas of  
non overlapping grains with :  
\begin{equation} 
S=-\int_V d{\vec r} ~ \{ {\rho\over\rho_0} \ln {\rho\over\rho_0} +  
(1-{\rho\over\rho_0})\ln (1-{\rho\over\rho_0}) \} 
\label{S} 
\end{equation} 
Here $\rho_0$ is the maximal accessible density. 
The minimum of the free energy gives 
the following Fermi-Dirac  
equilibrium density profile: 
\begin{equation} 
\rho({\vec r})=\rho_{FD}(z)=\frac{\rho_0}{e^{mg\rho_0\beta (z-\lambda)}+1} 
\label{FD} 
\end{equation} 
where $\lambda$ is a chemical potential which fixes the total number of 
particles. 
This result is in agreement with experimental \cite{JNBHM,RajCle} 
and other theoretical \cite{NCH,Haya-hong} findings.


The configurational temperature
can be evaluated
by introducing a small perturbation $\gamma U$ where $\gamma$ is 
a small parameter and U is an observable. If $U_r$ is the value of $U$ in the
inherent state $r$ the energy of the inherent state changes from $E_r$ to $E_r
+ \gamma U_r$. Using the fluctuation dissipation relation 
\begin{equation}
\left. \frac{\partial \lan U\ran }{\partial \gamma}\right|_{\gamma=0} 
= \beta(\lan U^2\ran - \lan U\ran^2) 
\label{fluctuation1}
\end{equation}
The ratio of the response function to the fluctuation of $U$ gives $\beta$. 
It is possible that such response function cannot be
measured experimentally. Another way to link $\beta$ to observable quantities
easier to measure, is the
following

From eq.(\ref{gibbs}) 
using standard statistical mechanics immediately follows 
\begin{equation}
\frac{\partial \lan E\ran }{\partial \beta} = \lan E^2\ran - \lan E\ran^2 
\label{fluctuation}
\end{equation}
which relates the fluctuation of the energy to the energy derivative.
The average is over the equilibrium distribution of eq.(\ref{gibbs}).  
Both the energy and its fluctuation, $C=\lan E^2\ran - \lan E\ran^2$, 
can be experimentally measured as function of $\Gamma$. 
Using  
$\frac{\partial \lan E\ran }{\partial \beta}=\frac{\partial \lan E\ran }{\partial \Gamma}\frac
{\partial \Gamma}{\partial \beta}$
from eq.~(\ref{fluctuation}) 
we obtain 
\begin{equation}
\beta(\Gamma) -\beta _0 = \int_{\Gamma_0}^{\Gamma} 
\frac{1}{C(\Gamma)}\frac{\partial \lan E\ran }{\partial \Gamma}d\Gamma 
\label{relation}
\end{equation}
where $\Gamma _0$ is a reference point which could be adequately chosen. 
Therefore a part from an additive
constant in principle it is possible to obtain $\beta$ as function of 
$\Gamma$. This procedure was experimentally proposed in ref.\cite{Knight}.  
We note that this approaches predicts an universal value of $\beta$, 
independent on the particular control parameter $\Gamma$ used in
different experiments\cite{nota1}.

So far for simplicity we have considered the
case in which the distribution is stationary.
From the experiments \cite{Knight} and from granular models \cite{NCH} 
we know that the dynamics often enters a quasi stationary
regime in which average macroscopic quantities, like the energy 
or density, changes extremely slowly in time.
In these situations we expect the distribution be quasi stationary
as in glass models\cite{sciortino}, namely 
still given by (\ref{gibbs}) where $\beta$ is slowly dependent on time. 
Here, $\beta$ can also be defined from generalized dynamical 
fluctuation dissipation 
relations (FDR) extended to granular media~\cite{N_fdt}. In analogy to 
eq.(\ref{fluctuation1}), in the linear response theory the FDR relates 
the system response to a perturbation to its time dependent 
correlation function in the unperturbed state \cite{FDR,FDR1}.


The distribution given by eq.(\ref{gibbs}) is the equilibrium distribution. 
When the system 
is not in equilibrium the probability $P_r$ will depend on the discrete
time $t_n = n\tau$ where $n$ is the $n$-th tap and  $\tau $ 
the duration of a single tap. 
In principle it is possible to write a master equation 
for such time dependent distribution:
\begin{equation} 
\frac{\partial P_r}{\partial t_n} = -\sum_s  P_r W_{rs} +  
\sum_r  P_s W_{sr} 
\label{ME} 
\end{equation} 
where $W_{rs}$ are 
the transition probabilities from the inherent state $r$ and the inherent
state $s$. If the system has to approach the equilibrium distribution
in the inherent states space
we can assume that detailed balance must be satisfied:
\begin{equation} 
W_{rs}/W_{sr}=\exp\left[-\beta(E_s-E_r)\right] ~ ~ .
\label{Wrs} 
\end{equation} 

Now we want to 
discuss how the statistical mechanics formalism discussed in this paper 
can work.
In a realistic model for granular medium, in order to 
reproduce a tap process and then calculate the macroscopic quantities,
one should  consider the detailed dynamical, including the energy
dissipated during the collision and then calculate the time average. 
The formalism developed here suggest  to calculate the 
macroscopic quantities when the system has reached a stationary distribution,
by taking ensemble averages instead of time averages.
Exchanging time averages with ensemble averages is in fact 
one of the powerful feature of statistical mechanics.
To construct the statistical ensemble one does not need to follow 
necessarily the exact dynamics but any dynamics which leads to stationarity.
In principle one could evaluate $P_r(t_n)$ directly from (\ref{Wrs}). 
Another possibility is to construct a dynamics which allows to visit
the inherent states and mimics the tapping process. 
For instance, at the simplest level, a single tap can be  
schematically reproduced by introducing an effective
time dependent bath temperature, $T_{bath}(t)$, which is a step function
which assumes two values: $T_{bath}(t)=T_{\Gamma}>0$ when the vibration is on
(and the grains average kinetic energy is larger than zero) and
$T_{bath}(t) = 0$ when the vibration is over (and the kinetic energy is zero).
A ``tapping sequence'' is the cyclical repetition of the above schematic tap 
\cite{NCH,our_rev}. 
This mimics the shaking process where $T_{\Gamma}$ plays the role of the
shaking amplitude $\Gamma$ in the experiments, and allows to explore the
inherent states visited when $T_{bath}=0$.
In general the parameter $T_{\Gamma}$ is different
from  the temperature $T_{conf}$ previously introduced to
describe stationary inherent states: $T_{\Gamma}$ is related to the
agitation of grains induced by the external drive, while $T_{conf}$
is the configurational temperature which characterizes the
``frozen'' inherent states. These two quantities are linked through
the functional relation (\ref{relation})  where $\Gamma$ is replaced
by $T_{\Gamma}$.

During the shaking process, when the kinetic energy is different from zero,
the state $k$ of the system is  not necessarily a mechanical stable
configuration.
Say ${\cal P}_k(t)$ the probability that the system is in one of such
states at time $t$.
After each tap the system goes into one of the inherent sates, say $r$.
Thus the probability, $P_r(t_n)$, to be in the inherent state $r$
after the $n$-th tap is related to the the probability
${\cal P}_r(t)$ by the following relation:
\begin{equation}
P_r(t_n)={\cal P}_r(n\tau)
\end{equation}
So, within the above models it is possible to evaluate the time dependent
distribution function for such particular dynamics and the corresponding
energy. At stationarity say E the value of the average energy,
the asymptotic distribution 
is given by 
eq.(\ref{beta}) with $\beta$ corresponding to the energy E.  
The above simplified approach to the dynamics of granular media has been
recently also discussed in Ref. \cite{brey}, with an interesting
detailed discussion.

Like in statistical mechanics  the basic postulate that time
averages can be substituted by Gibbs ensemble averages 
can be considered as a working hypothesis which 
cannot be proven but can be accepted only a posteriori if the results 
which follow are correct. 

\begin{figure}[ht]
\centerline{\hspace{-2cm} 
\psfig{figure=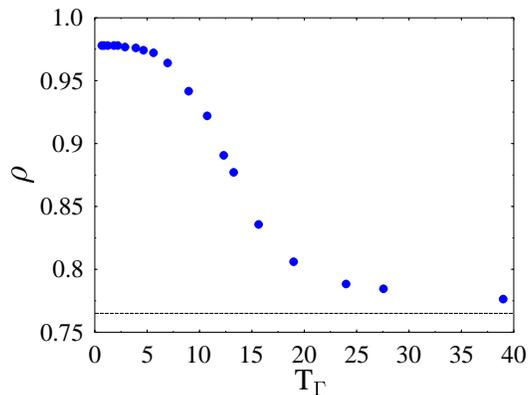,height=8cm,angle=-90}}
\vspace{-1.5cm}
\caption{The average density of the inherent states, $\rho$, 
of a model of granular media (Tetris) as a function of the adimensional 
shaking amplitude $T_{\Gamma}$. 
The asymptote (the horizontal dashed line) corresponds
to the minimal stability density $\rho_m$. 
} 
\label{dens_inh}
\end{figure}

There are many lattice models  for granular media 
\cite{CH,NCH,our_rev} which are based on the above tap dynamics.
They are based on an Hamiltonian formalism which takes into account 
gravity and hard core repulsions which may generate geometric frustration 
by not allowing two grains to overlap (for a review see \cite{our_rev}). 
Although the models are rather crude, they 
have given general results in excellent agreement with 
experimental findings ranging from logarithmic compaction to segregation,
density fluctuations, 
``memory'' and ``aging'' effects, Fermi-Dirac density profiles, 
non Gaussian density distributions and others 
(for a review see \cite{our_rev}).  
As an example, we show in Fig.\ref{dens_inh} the average density of the
inherent states, $\rho$, in one of these models, 
the Tetris \cite{our_rev}, as a function of $T_{\Gamma}$. 
The system is ``shaken'' 
with a given $T_{\Gamma}$ for a long time and, after switching off the 
shaking, the density of the obtained ``frozen'' state is recorded. 

It is very interesting that 
the results of Fig.\ref{dens_inh} reproduce qualitatively the
reversible line in granular compaction \cite{Knight} 
and closely remember the energy of inherent structures 
as a function of their equilibration
temperature before quenching \cite{Stillinger}.

In conclusion, in this paper we have elaborated the original
Edwards' statistical mechanics  approach to  
granular media to describe the quasi-static distribution  
of the system in their state at rest that we have called inherent states,
in analogy with glassy systems. We have distinguished two different 
temperatures the zero external bath temperature and the internal
configuration temperature related to Edwards' compactivity, 
which can be experimentally 
measured and linked to the driving $\Gamma$.
We have postulated  a Master equation, eq.(\ref{ME}), for the 
evolution of granular packs among its  inherent 
states.  
A straightforward application of these ideas reproduces the Fermi-Dirac density
distribution found experimentally.

Models from standard statistical mechanics, which can be well understood in  
the present unifying formalism, are able to describe a variety of properties 
of granular media  
\cite{Edwards,mehta1,CH,NCH,our_rev,Haya-hong,N_fdt,brey,head}. 

This work was partially supported by the TMR Network ERBFMRXCT980183, 
INFM-PRA(HOP) and INFM Parallel Computing Initiative.

\vspace{-1.8cm}


\begin{thebibliography}{40}
\bibitem{nota1} 
For sake of brevity, here we use the symbol $\Gamma$ to represent 
all the set of relevant control parameters.
\bibitem{nota_qs} Even if quasi-equilibrium is reached our basic assumption 
is legitimated, like the physics of glassy systems suggests.
\end{thebibliography}
\end{document}